\documentclass[11pt,journal]{IEEEtran}   
\usepackage{amsmath,amssymb,amsfonts}
\usepackage{bbm}
\usepackage{bm}
\usepackage{comment}
\usepackage{amsthm}
\usepackage[caption=false,font=normalsize,labelfont=sf,textfont=sf]{subfig}
\usepackage{algorithm}
\usepackage{algpseudocode}
\usepackage{array}
\usepackage{textcomp}
\usepackage{stfloats}
\usepackage{url}
\usepackage{verbatim}
\usepackage{graphicx}
\usepackage{cite}
\usepackage{color}

\AtBeginEnvironment{quote}{\it}

\usepackage{tikz}
\usepackage{pgfplots}
\usepackage{tikzscale}
\usepackage{tikz-qtree}
\usepgfplotslibrary{fillbetween}
\usetikzlibrary{plotmarks}
\usetikzlibrary{patterns}
\usetikzlibrary{positioning}
\usetikzlibrary{arrows}
\usetikzlibrary{shapes}
\usepackage{multirow}
\usepackage{xcolor}
\usepackage{booktabs}
\usepackage{glossaries}

\newacronym{mse}{MSE}{mean square error}
\newacronym{mmse}{MMSE}{minimum mean square estimator}
\newacronym{ml}{ML}{machine learning}
\newacronym{ai}{AI}{artificial intelligence}
\newacronym{urllc}{URLLC}{ultra-reliable low-latency communications}
\newacronym{arq}{ARQ}{automatic repeat request}
\newacronym{qos}{QoS}{quality of service}
\newacronym{pdf}{PDF}{probability density function}
\newacronym{pmf}{PMF}{probability mass function}
\newacronym{iid}{i.i.d.}{independent and identically distributed}
\newacronym{cdf}{CDF}{cumulative density function}
\newacronym{iot}{IoT}{Internet of Things}
\newacronym{aoi}{AoI}{age of information}
\newacronym{ldpc}{LDPC}{low-density parity check}
\newacronym{jscc}{JSCC}{joint source-channel coding}
\newacronym{paoi}{PAoI}{peak age of information}
\newacronym{snr}{SNR}{signal-to-noise ratio}
\newacronym{voi}{VoI}{value of information}
\newacronym{kpi}{KPI}{key performance indicator}
\newacronym{ccdf}{CCDF}{complementary CDF}
\newacronym{gm}{GM}{Gauss-Markov}
\newacronym{dnn}{DNN}{deep neural network}
\newacronym{cnn}{CNN}{convolutional neural network}
\newacronym{MDP}{MDP}{Markov decision process}
\newacronym{mssim}{MS-SSIM}{multi-scale structural similarity index measure}
\newacronym{MA-POMDP}{MA-POMDP}{multi-agent partially observable Markov decision process}
\newacronym{e2e}{E2E}{end-to-end}
\newacronym{ntn}{NTN}{non-terrestrial networks}
\newacronym{feel}{FEEL}{federated edge learning}
\newacronym{aiot}{AIoT}{AI of Things}
\newacronym{mvcnn}{MVCNN}{multi-view CNN}
\newacronym{c2}{C2}{joint communication-and-computing}
\newacronym{ps}{PS}{parameter server}
\newacronym{pa}{PA}{perfect accumulation}
\newacronym{uma}{UMA}{unsourced massive access}
\newacronym{obda}{OBDA}{one-bit digital accumulation}
\newacronym{gdoac}{GD-AirComp}{generalized digital AirComp}

\definecolor{darkblue}{HTML}{00264D}
\definecolor{darkgreen}{HTML}{0F4233}
\definecolor{darkred}{HTML}{800000}
\definecolor{darkgrey}{HTML}{303030}

\definecolor{color1}{HTML}{0000FF}
\definecolor{color2}{HTML}{A627CF}
\definecolor{color3}{HTML}{DB57A2}
\definecolor{color4}{HTML}{FA8775}

\definecolor{lightgreen}{HTML}{42FF42}
\definecolor{lightblue}{HTML}{4242FF}
\definecolor{lightred}{HTML}{FF4242}
\definecolor{lightpurple}{HTML}{CF00CF}

\newcommand{\edit}[1]{#1}

\graphicspath{{./Figures/}}

\begin{document}
\title{Timely and Massive Communication in 6G: Pragmatics, Learning, and Inference}

\author{Deniz G{\"u}nd{\"u}z,~\IEEEmembership{Fellow,~IEEE}, Federico Chiariotti,~\IEEEmembership{Member,~IEEE}, Kaibin Huang,~\IEEEmembership{Fellow,~IEEE}, Anders E. Kal{\o}r,~\IEEEmembership{Member,~IEEE}, Szymon Kobus~\IEEEmembership{Student Member,~IEEE}, and Petar Popovski,~\IEEEmembership{Fellow,~IEEE}\\
}

\maketitle

\begin{abstract}
5G has expanded the traditional focus of wireless  systems to embrace two new connectivity types: ultra-reliable low latency and massive communication. The technology context at the dawn of 6G is different from the past one for 5G, primarily due to the growing intelligence at the communicating nodes. This has driven the set of relevant communication problems beyond reliable transmission towards semantic and pragmatic communication. This paper puts the evolution of low-latency and massive communication towards 6G in the perspective of these new developments. At first, semantic/pragmatic communication problems are presented by drawing parallels to linguistics. We elaborate upon the relation of semantic communication to the information-theoretic problems of source/channel coding, while generalized real-time communication is put in the context of cyber-physical systems and real-time inference. The evolution of massive access towards massive closed-loop communication is elaborated upon, enabling interactive communication, learning, and cooperation among wireless sensors and actuators.  
\end{abstract}
\glsresetall
 
\begin{IEEEkeywords}
6G, AI-based networking, massive access, machine-to-machine communications, pragmatic communication, semantic communication
 \end{IEEEkeywords}

\section{Introduction}

The focus of wireless cellular evolution before 5G was on high-fidelity delivery of voice and reliable data transmission at increasing rates. 4G can be seen as a reliable broadband data pipe, geared towards human-operated mobile devices. 5G expanded the wireless landscape by considering autonomous machine-type devices, robots, as well as a new class of tactile, human-operated devices. This led to two new connectivity types in 5G: low latency, coupled with high reliability, and massive \gls{iot} communication. 
Tracing the further evolution of timely and massive communication needs to account for the 6G features that came to prominence via various technology visions. The premises in this article are: 

\begin{itemize}
 \item With the advances in hardware technology and \gls{ml} and \gls{ai} algorithms, the intelligence available to communicating devices is steadily increasing, leading to potential synergies between communication, computation, and learning. Transmitted data can be used for the learning process and, vice versa, communication can be aided by the intelligence at the nodes and embedded in the protocols. As the vision of networked intelligence is emerging, 
 the problem of communication as an accurate reproduction of data bits needs to be generalized to the problem of communicating the relevant information for the desired task, which can include computation, learning and inference, summarized in the phrase \emph{semantic communication}~\cite{popovski2020semantic, lan2021semantic, gunduz2022beyond}.

 \item Recently, there has been a growing awareness that low latency is just one example of a variety of timing constraints in a communication system~\cite{popovski2022perspective}. In practice, the timing requirements belong to a more general category in which the usefulness of communication is gauged through a goal accomplishment. This is the basis for \textit{pragmatic communication}, \edit{often called goal-oriented or effective communication in the literature,} where cyber-physical systems interact with each other and the environment.   


 \item  Massive IoT communication has been mainly considered as a mechanism to offload data collected by edge nodes to a central or Cloud server. Thus, massive access has traditionally been defined as a problem for uplink transmission. Nevertheless, new applications call for revising this assumption, as various real-time systems require closed-loop communication with a massive number of sensors and actuators. The data acquired by IoT devices can be used for distributed training of \gls{ml} models, using both uplink and downlink communications. 
\end{itemize}

Based on these observations, this paper will describe the evolution of 6G systems towards a \emph{network of intelligence} augmenting low-latency and massive communications considered in 5G. We start with a discussion on semantic and pragmatic communications by drawing parallels to linguistics. Next, we describe the relation between semantic communication and data/knowledge representation seen through the lens of source coding. This is followed by generalized real-time requirements in 6G, including both cyber-physical systems that interact with the environment
and real-time inference. Finally, we discuss closed-loop massive communication and its potential role in distributed and federated learning. 

\section{Semantic and Pragmatic Communication}

A standard introduction to semantic communication refers to the preface that W. Weaver wrote in~\cite{ShannonWeaver49} to the original work of C. Shannon from 1948. Weaver defined three types of communication problems: \emph{(Level A)} The technical problem: accurate transmission of bits from the transmitter to the receiver; \emph{(Level B)} The semantic problem: how precisely do the transmitted symbols convey the desired meaning; and \emph{(Level C)} The effectiveness problem: how the received meaning affects conduct in the desired way. Despite the initial intuitive appeal of this classification, the distinction between semantics and effectiveness and the corresponding communication models have remained blurry. This is also reflected in the often used phrase ``semantic and goal-oriented communications,'' which treats the beyond-technical problems in a bulk, avoiding to make a clear distinction; yet, the term ``goal-oriented'' presumably covers the part of effectiveness. 

\begin{figure}[t]
    \centering
    \begin{tikzpicture}[every text node part/.style={align=center}]

\node[rectangle,draw,darkblue,fill=darkblue,name=ling,minimum height=1cm,minimum width=2cm,text width=1.7cm] at (0,0) {\footnotesize\textcolor{white}{\textbf{Linguistic domains}}};
\node[rectangle,draw,darkblue,fill=white!80!darkblue,name=prag,minimum height=1cm,minimum width=2cm,below=0cm of ling] {\footnotesize\textcolor{darkblue}{Pragmatics}};
\node[rectangle,draw,darkblue,fill=white!80!darkblue,name=sem,minimum height=1cm,minimum width=2cm,below=0cm of prag] {\footnotesize\textcolor{darkblue}{Semantics}};
\node[rectangle,draw,darkblue,fill=white!90!darkblue,name=synt,minimum height=1cm,minimum width=2cm,below=0.2cm of sem] {\footnotesize\textcolor{darkblue}{Syntax}};
\node[rectangle,draw,darkblue,fill=white!90!darkblue,name=morp,minimum height=1cm,minimum width=2cm,below=0cm of synt] {\footnotesize\textcolor{darkblue}{Morphology}};
\node[rectangle,draw,darkblue,fill=white!90!darkblue,name=phon,minimum height=1cm,minimum width=2cm,below=0cm of morp] {\footnotesize\textcolor{darkblue}{Phonology}};

\node[rectangle,draw,darkgreen,fill=darkgreen,name=comm,minimum height=1cm,minimum width=2.2cm,text width=2.2cm,right=0.6cm of ling] {\footnotesize\textcolor{white}{\textbf{Communication systems}}};
\node[rectangle,draw,darkgreen,fill=white!80!darkgreen,name=semcom,minimum height=2cm,minimum width=2.2cm,text width=2.2cm,below=0cm of comm] {\footnotesize\textcolor{darkgreen}{Semantic and goal-oriented communications}};
\node[ellipse,draw,darkgreen,fill=white!90!darkgreen,name=cnt,minimum height=0.8cm,minimum width=2cm,text width=1.5cm,right=0.6cm of synt] {};
\node[ellipse,draw,darkgreen,fill=white!90!darkgreen,name=pkt,minimum height=0.8cm,minimum width=2cm,text width=1.5cm,right=0.6cm of morp] {};
\node[ellipse,draw,darkgreen,fill=white!90!darkgreen,name=phy,minimum height=0.8cm,minimum width=2cm,text width=1.5cm,right=0.6cm of phon] {{\tiny\textcolor{darkgreen}{PHY symbols}}};

\node[name=pkt1] at (2.85,-4.15) {{\tiny\textcolor{darkgreen}{Composite symbols}}};
\node[name=pkt2,below=-0.2cm of pkt1] {{\tiny\textcolor{darkgreen}{(packets)}}};
\node[name=cnt1] at (2.85,-3.1) {{\tiny\textcolor{darkgreen}{Error control,}}};
\node[name=cnt2,below=-0.2cm of cnt1] {{\tiny\textcolor{darkgreen}{protocol verification}}};

\node[rectangle,draw,darkred,fill=darkred,name=sha,minimum height=1cm,minimum width=2.2cm,text width=2.2cm,right=0.6cm of comm] {\footnotesize\textcolor{white}{\textbf{Shannon-Weaver}}};
\node[rectangle,draw,darkred,fill=white!80!darkred,name=levc,minimum height=1cm,minimum width=2.2cm,text width=2.2cm,below=0cm of sha] {\footnotesize\textcolor{darkred}{Level C} \tiny{\textcolor{darkred}{Effectiveness problem}}};
\node[rectangle,draw,darkred,fill=white!80!darkred,name=levb,minimum height=1cm,minimum width=2.2cm,text width=2.2cm,below=0cm of levc] {\footnotesize\textcolor{darkred}{Level B} \tiny{\textcolor{darkred}{Semantic problem}}};
\node[rectangle,draw,darkred,fill=white!90!darkred,name=leva,minimum height=3cm,minimum width=2.2cm,text width=2.2cm,below=0.2cm of levb] {\footnotesize\textcolor{darkred}{  Level A  } \tiny{\textcolor{darkred}{Technical problem: correct transmission}}};

\draw[-,dashed,darkgrey] (prag.east) -- ([yshift=0.5cm]semcom.west);
\draw[-,dashed,darkgrey] (sem.east) -- ([yshift=-0.5cm]semcom.west);
\draw[-,dashed,darkgrey] (synt.east) -- (cnt.west);
\draw[-,dashed,darkgrey] (morp.east) -- (pkt.west);
\draw[-,dashed,darkgrey] (phon.east) -- (phy.west);
\draw[-,dashed,darkgrey] (levc.west) -- ([yshift=0.5cm]semcom.east);
\draw[-,dashed,darkgrey] (levb.west) -- ([yshift=-0.5cm]semcom.east);

\draw[-,dashed,darkgrey] (cnt.east) -- (leva.west);
\draw[-,dashed,darkgrey] (pkt.east) -- (leva.west);
\draw[-,dashed,darkgrey] (phy.east) -- (leva.west);

\draw[-,dotted,darkgrey,thick] ([xshift=-0.2cm,yshift=-0.1cm]sem.south west) -- ([xshift=0.2cm,yshift=-0.1cm]levb.south east);

\end{tikzpicture}
    \caption{The five linguistic domains, \edit{their relation to different elements of communication engineering} and Weaver's classification.}
    \label{fig:linguistic}
    \vspace{-0.6cm}
\end{figure}
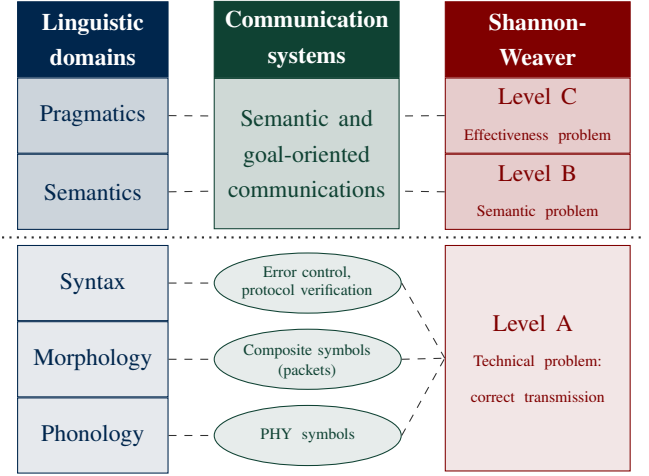

As an attempt for a clearer distinction between Levels B and C, we resort to the classification in linguistics~\cite{gleason2022development}, as shown in Fig.~\ref{fig:linguistic}, and relate them to problems in communication engineering, as well as to Weaver's classification. \emph{Phonology} deals with the speech sounds and phonemes as basic ingredients of a language. \edit{A phoneme can be related to a physical layer symbol from a given constellation used in  communication engineering or a channel use in an information-theoretic model.} \emph{Morphology} studies the minimal meaningful units of language called morphemes. \edit{In a communication system, such a meaningful unit can be a group of symbols, such as an OFDM symbol, a codeword, or a packet.} \emph{Syntax} deals with the rules to form sentences and can thus be related to the correct protocol operation. \edit{Some syntactic elements are also present in individual packets, as they are data units with integrity check that allow design of link-layer protocols with error detection}. Regardless of that, the first three linguistic domains can be mapped, as a group, to \edit{various elements that constitute the technical communication problem, as illustrated in Fig.~\ref{fig:linguistic}.} 
\emph{Semantics} studies the signs and their relation to the world\footnote{\edit{Following the argument of C. W. Morris on the ambiguity of ``meaning'' due to overuse, we have adopted his definition of semantics, without  the word ``meaning''. See: C. W. Morris, Foundations of the Theory of Signs. University of Chicago Press, 1938.}}.
Finally, \emph{pragmatics} refers to the rules in conversation and social contexts, including implicit communication through external cues and shared context. 
\edit{Fig.~\ref{fig:linguistic} shows that these two domains,  corresponding to Weaver's semantic and effectiveness levels, are not considered in the design of modern communication systems, which focus on the technical problem, and therefore, has motivated a flurry of recent research activity.} 
Against this background, we will first attempt to define a distinction between semantic and pragmatic communications. \edit{Our choice to use the term ``pragmatic'' to refer to Level C problems is motivated by a closer match with the linguistic metaphor, which clarifies the distinction between the two.}

\emph{Semantic communication} refers to the the unidirectional transmission of a message or a signal $x$ under a prescribed measure of quality that relies on the semantics, along with the utilized source/channel coding. As such, it does not need to be related to the physical time, but can work with an abstract communication model, consisting of channel uses. Time is not explicitly a part of Shannon's mathematical model of communication; that model provides a causal relation between inputs and outputs, but says nothing about the time duration between two inputs, two outputs, or between an input and its corresponding output. Semantic communication can be put in a temporal context, such as ``convey a certain meaning within a given time interval $T$'' by, e. g., establishing a symbol duration $T_s$, a codeword duration $T$ and codeword length of $n=\frac{T}{T_s}$ symbols.

\emph{Pragmatic communication} captures the interaction between communicating entities, machines or humans, as well as interaction with their physical environment, such as movement or actuated command, which determines the dynamic context. Dealing with the physical time is thus necessary in pragmatic communication, as the interaction between the entities via communication and/or action occurs under the constraints of a physical time. Note that pragmatic communication is not identical to multi-user communication, as there are multi-user channels in which the transmitting users are not interacting with each other, such as the multiple access or broadcast channels, which can also be seen as setups for unidirectional semantic communication.


\section{Semantic Communication}\label{s:semantics}

Semantic communication has recently become a catchphrase to describe communication systems that cater for the content of the source signal rather than focusing solely on the reliable transmission of bits. Current communication protocols are designed to create the largest capacity reliable bit pipes with minimal resources, i.e., maximize the capacity in terms of bits per signal dimension, while achieving the highest reliability with the highest energy efficiency. In that sense, the codes and modulation schemes employed at the physical layer are independent of the delivered content and transfer the packets of bits to their destinations. Content is handled at the application layer, where different types of information sources, e.g., text, image, or video, are transformed into packets of bits, with the goal of obtaining the most efficient representation that allows the receiver to reconstruct the information with the desired fidelity. Thus, the semantic aspects of the source are currently handled at the application layer. 

{Semantic communication is commonly exemplified through} the communication of a text or an image. Semantic-aware communication conveys the meaning of a text without sending the exact order of words. For images, semantics typically refers the objects in the image and their relative locations, without necessarily reconstructing them with a pixel-level fidelity. Extracting the meaning of a sentence or the semantic content of an image are highly challenging tasks, but data-driven \gls{ml} algorithms has led to significant progress. {Despite the impression that this type of semantic communication departs from the basic information theoretic principles,} these problems can still be treated within the framework of rate-distortion theory, dealing with an efficient representation of the source signal within the desired fidelity. However, unlike the problems that one encounters in a first course on rate-distortion theory, many practical sources do not have \gls{iid} samples, and we typically do not have a well-defined additive fidelity measure, such as \gls{mse}. Indeed, a major challenge is to identify the fidelity measure itself, even before specifying a compression algorithm. Shannon was well-aware of the limitations of the assumptions in his theory, but these were necessary to obtain the succinct single-letter mathematical expressions that connect source compression with channel coding through the mutual information functional. However, the complexity of this setup can be addressed through \gls{ml} algorithms. This approach to semantic communications that relies on generalized distortion measures has become relevant with the recent advances in \gls{ml} algorithms and the intelligence of communicating devices.

In his seminal paper~\cite{ShannonWeaver49}, after introducing the fidelity measure $\rho(x,y)$ between the input signal $x$ and its reconstruction $y$, Shannon adds the following remark: 

\smallskip

\noindent \emph{``The structure of the ear and brain determine implicitly a number of evaluations, appropriate in the case of speech or music transmission. There is, for example, an `intelligibility' criterion in which $\rho(x,y)$ is equal to the relative frequency of incorrectly interpreted words when message $x(t)$ is received as $y(t)$. Although we cannot give an explicit representation of $\rho(x,y)$ in these cases, it could, in principle, be determined by sufficient experimentation. Some of its properties follow from
well-known experimental results in hearing, e.g., the ear is relatively insensitive to phase and the sensitivity to amplitude and frequency is roughly logarithmic.''}

\smallskip

One can argue that, here, Shannon makes an early argument for a data-driven approach to measuring fidelity, as done in state-of-the-art \gls{ml} algorithms.  

It is possible to extend this framework to include other types of fidelity measures. For example, rather than reconstruction, the receiver may want to only estimate certain statistics of the source signal; or infer another random variable correlated with the source signal, which can correspond to, e.g. the class of the input sample. \edit{Most of these problems can be considered in the context of remote source coding~\cite{Dobrushin}: the encoder has access to correlated observation $z$ with the underlying source $x$, which the receiver wants to reconstruct.} For example, $z$ may represent the class the input $x$ belongs to, or its features. In~\cite{Dobrushin}, $x$ and $z$ come from a known joint distribution $p(x,z)$, and i.i.d. samples $z^n=(z_1, \ldots, z_n)$ of $z$ are available at the encoder, while the decoder wants to decode $x^n$. This is a  generalization of Shannon's rate-distortion problem, reducible to Shannon's original formulation by an appropriate change of the distortion measure. This formulation is quite general, and can cover many statistical problems under communication constraints. For example, when the fidelity measure between the reconstruction $y^n$ and the source sequence $x^n$ is the \textit{log-loss}, we can obtain a single-letter expression for the remote rate distortion function, equivalent to the well-known \textit{information bottleneck} function. 

\subsection{Joint Source-Channel Coding (JSCC)}\label{ss:JSCC}

Separation of source and channel coding leads to an architecture that is modular {and universal}, where bits coming from different sources can all be delivered using the same channel code. This simplifies the code design for each subproblem, for example, code design for compressing different sources, or reliable communication over different types of channels, and resulted in the specialization of engineers and researchers to specific subproblems. The separate architecture also allows encryption straightforwardly: as the communication system is oblivious to the {data source}, the {random bits obtained as a result of the compression} can be encrypted with no impact on {the way they} can be transmitted. The theoretical basis for the separated architecture is Shannon's separation theorem~\cite{ShannonWeaver49}, showing that separation is asymptotically optimal for ergodic sources and channels. 

Nevertheless, there are many scenarios in which a joint design can provide performance gains and simplify the code design at the expense of modularity. A prominent example is the transmission of an \gls{iid} source sequence of Gaussian samples, $x^n$, over $n$ uses of an additive white Gaussian channel under the \gls{mse} distortion measure. Here, the optimal scheme consists of transmission of each source sample over one channel use, by simple scaling according to the transmitter power constraint, and estimating each sample with a \gls{mmse} at the decoder. This simple uncoded scheme offers the lowest possible latency, while  achieving the same average \gls{mse} performance for any $n$ that can be achieved by the separation scheme as $n \rightarrow \infty$. If we instead consider two receivers with different \glspl{snr}, the uncoded scheme simultaneously achieves the optimal performance for both receivers as if each one is the sole receiver in the system; while this is not possible with standard channel codes that operate at the capacity region of the underlying broadcast channel. Yet, until recently, no practical \gls{jscc} design provided significant gains compared to separation based codes for practical sources such as images or videos. Most of the existing designs relied on separate source and channel codes, whose parameters are optimized jointly in a cross-layer fashion. While this approach retained some modularity, the gains are mainly through adapting the source/channel code parameters according to temporal variations of the underlying source and channel statistics. Despite certain performance gains are achieved for non-ergodic channels/sources, they are not truly joint code designs.

{More recently, significant progress has been made by employing deep learning in \gls{jscc} problems. 
Pioneered {in} \cite{Eirina:TCCN:19}, these truly joint designs are known as DeepJSCC, where the encoder and decoder are parameterized as a pair of \glspl{dnn}, trained jointly on a given source dataset and channel distribution. DeepJSCC provides several significant advantages: First of all, even for a given fixed channel \gls{snr}, for which the right compression and channel code rates can be chosen to achieve reliable communication with high probability, DeepJSCC can achieve a better performance compared to the concatenation of state-of-the-art compression codecs (e.g., BPG or JPEG200) and channel codes, e.g., \gls{ldpc}. The results show that the gains of DeepJSCC are larger for low bandwidth ratio (average channel uses per source sample) and low \gls{snr} regimes. The gains also become more pronounced when a perceptually aligned distortion measure, such as \gls{mssim}, is considered, or the model is trained and tested on data from a specific domain, such as satellite images. This is because DeepJSCC can be trained with a specific loss function or a dataset, while conventional compression algorithms are not adaptive. Moreover, DeepJSCC simplifies the coding process, and can provide significant reduction in complexity. The architecture in \cite{Eirina:TCCN:19} is a simple 5-layer \gls{cnn}, which can be parallelized and implemented efficiently on a dedicated hardware, whereas conventional compression codecs and iterative decoding algorithms are computationally much more demanding. On the other hand, the original design in \cite{Eirina:TCCN:19} requires training a separate encoder-decoder network pair for each channel condition (\gls{snr}/bandwidth ratio), which limits its practicality as it would require storing different network parameters to be used in different scenarios. This would impose significant memory complexity. Several adaptive architectures have been subsequently proposed, showing that a single DeepJSCC network can be deployed for a range of channel and bandwidth values \cite{Kurka:TWC:21}.

\edit{Learning-based solutions to semantic communication bring about another unexplored dimension of compression and \gls{jscc} problems. In conventional information theoretic approaches to these problems, we assume perfectly known source and channel statistics, and provide certain performance guarantees under the ergodicity assumption. Alternatively, in universal schemes, the distribution is either learned from a first pass over the particular input sequence to be compressed, or the codebook is adapted dynamically to the sequence as it is being coded, as in the well-known Lempel-Ziv algorithm. In the aforementioned ML approaches to \gls{jscc}, we assumed an offline training stage that learns the code using a specific dataset of images and a channel distribution. On the other hand, if the source statistic changes over time, an online learning approach can be adapted, in which case the overhead of communicating the updated model should also be taken into account.}


\section{Generalized 6G Real-time Communication}\label{s:real-time}


This section treats the role of timing in communication systems, which is an essential component of the engineering formulation of pragmatic communications. We will explain how it generalizes the technical communication problem and highlight its connections to other timing-based performance measures. 

\subsection{Pragmatic Communication}

Formulating a communication problem of effectiveness/pragmatics (Level C) and favorably affecting an agent's conduct requires a quantifiable way to measure the effect of its actions. A general way to model series of actions and their outcomes is a \gls{MDP}, where an agent acts in a stochastic environment and aims to accumulate as much reward as possible. Specifically, the current state of the environment and the action of the agent affect the probability of the consequent state and the reward. 

The effectiveness/pragmatic communication problem is formulated in \cite{Tung;JSAC:21} as an extension of the \gls{MDP} framework to a \gls{MA-POMDP}, in which two agents communicate to maximize their common reward. The agents represent different parties, such as transmitters, receivers, controllers, and robots, and partial observability means that some information is not known to all the parties in the system. The joint actions of the agents influence the evolution of the environment. A simple setup that captures the fundamental aspects is a \emph{remote \gls{MDP}}, as formulated in \cite{Tung;JSAC:21} and shown in Fig.~\ref{f:remoteMDP}. Here a controller, say a guide, observes the environment and transmits messages through a noisy channel to instruct and navigate the agent towards a target location. The actions of the agent can be possible movements in the environment, while the actions of the guide are the messages it transmits. A large positive reward is received when the agent reaches the target, while a unit of negative reward is accrued at each time. 
This incentivizes to reach the target as soon as possible, as the goal is to maximize the total cumulative reward. Note that the timing here is in reference to the interactions with the environment, while the channel can still be used multiple times for the transmission of each message. Differently from Level A or B problems, here the communication performance cannot be measured by the reliability of bits transmitted at each round, or the reconstruction quality of the receiver. What matters is the set of actions taken over the horizon until the target is reached, while maximizing the cumulative reward, and the context/state of the receiver is critical to determine the success/quality of each transmission, as in the definition of pragmatic communication in linguistics.

Moreover, this formulation includes Level A and B problems as special cases \cite{Tung;JSAC:21}. If the state of the environment is a randomly generated bit sequence, and the goal of the agent is to match this bit sequence, then we recover the classical channel coding problem, where maximizing the average reward over a fixed time horizon is equivalent to minimizing the error probability of a fixed-length code. If the state of the environment is a randomly generated sequence of \gls{iid} symbols from a fixed distribution $p(x)$ and the goal of the agent is to match this sequence with the minimum distortion, then we recover the lossy source coding, or the semantic communication problem as formulated in Section \ref{s:semantics}. 

The notion of pragmatic communication can be extended to source coding \cite{Kobus:ISIT:23}, where the standard objective is to minimize the expected length of the description of a random symbol drawn from an alphabet. To pose it as an effectiveness problem, let the alphabet represent the set of possible states the receiver can occupy: the aim is to communicate a randomly drawn state within the minimum average time; however, as in a remote \gls{MDP}, the receiver's state cannot change arbitrarily but there is a constrained set of possible transitions. This setting can be interpreted as guiding an agent through a graph of possible states from the initial to the goal state. At each time step, the agent receives a message and decides a state transition. The simplest strategy to communicate such a plan is to code each state transition sequentially; this has the benefit of immediate relevance to the controlled agent. The opposite approach is to only send the goal, which is more efficient in terms of total communicated information, but introduces a greater delay before actionable information. In general, a right balance needs to be achieved between prioritizing actionable information in the short term and information useful over a long term, bringing timing aspects into a simple compression problem.

\begin{figure}[t]
    \centering
    \input{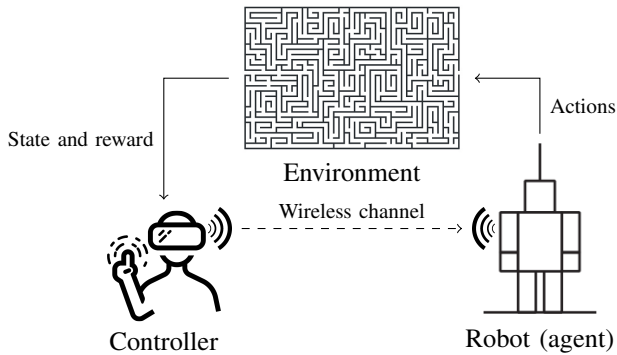}
     \caption{Illustration of the remote MDP problem exemplifying level C pragmatic communications: an intelligent controller controls an agent over a wireless channel to maximize its reward through interactions with the environment.}
    \label{f:remoteMDP}
    \vspace{-0.6cm}
\end{figure}


\subsection{Timing Constraints and Pragmatic Communication} 

The underlying assumption behind defining timing-related QoS  is that the delay between the generation of a piece of information and its reception and processing by the receiver affects the control performance~\cite{popovski2022perspective}. \edit{The nature of this assumption becomes clear for Level C problems:} timing is useful because information about a process becomes less accurate over time, as the real environment drifts from the observed state.

Latency and \gls{aoi} are textbook Level A metrics~\cite{popovski2022perspective}, as they do not consider the content of updates, only how quickly they are delivered (latency) or how stale the information available at the receiver gets (\gls{aoi}). If we consider a tracked signal $x(t)$, which is observed by the transmitter, the latency of packet $i$, generated at time $g_i$ and received at time $r_i$, is simply $\tau_i=r_i-g_i$, while the \gls{aoi} at time $t>r_i$ (but before any other, more recent, packet is received) is $\Delta(t)=t-g_i$. We can see that \gls{aoi} $\Delta(t)$ is tied more closely to the tracked process than the latency: as the last update from the transmitter has an age $\Delta(t)$, the receiver must estimate the current value of the process, $\hat{x}(t)$, using only the historic data of the signal up until time $g_i$.

The \gls{voi} considers the \emph{content} of messages and not just their timing~\cite{alawad2022value}. If the receiver has a known estimator, the value $\nu(t)$ of an update generated at time $t$ is given by the difference its transmission would make in the error. The receiver gets a series of observations $h(t)=(y_1,g_1),\ldots,(y_i,g_i)$, which correspond to the packets transmitted up until time $t$ by the sensor observing the process and their generation instants. The observations may be noisy, requiring further processing. The receiver keeps an estimate $\hat{x}(t|h(t))$ and the error function can be defined as $e(x(t),\hat{x}(t|h(t)))$. The \gls{voi} of an update is defined as:
\begin{equation}\label{eq:voi}
\begin{aligned}
\nu(t)=&e(x(t),\hat{x}(t|h(t)))-e(x(t),\hat{x}(t|h(t),y(t)))
\end{aligned}
\end{equation}
For well-defined $e(x(t),\hat{x}(t|h(t)))$ (e.g.~$\ell_2$ norm), and the receiver has a good estimator, \gls{voi} is always positive, as additional information will never increase the error.

We can further distinguish between \emph{push-based} communication systems, where the sender decides autonomously when to send data, and \emph{pull-based} systems, where the sender only sends updates in response to requests from the receiver. In the pull-based case, the \gls{voi} definition in~\eqref{eq:voi} cannot use the value of $y(g_i)$, as the receiver does not know it before it is transmitted, but must estimate the \gls{voi} as well, based on its current knowledge~\cite{holm2023goal}:
\begin{equation}\label{eq:voi_rx}
\begin{aligned}
\nu'(t)=&\mathbb{E}\left[e(x(t),\hat{x}(t)|h(t))\right]\\
&-\mathbb{E}\left[e(x(t),\hat{x}(t|h(t),y(t))|h(t)\right].
\end{aligned}
\end{equation}
If the process is \emph{memoryless}, i.e., the residual error of the estimator is a martingale (e.g., a Wiener process), the Level B problem of \gls{voi} maximization reduces to a simpler Level A \gls{aoi} minimization. In this case, the estimate of the current state can be performed knowing only $\Delta(t)$ with the same precision as the estimate knowing the full history $h(t)$, and minimizing a (potentially non-linear) function of $\Delta(t)$ is equivalent to maximizing the receiver-computed \gls{voi}. We can contrast this with \emph{stateful} processes, which include most Markov processes, where the value of $x(t)$ affects future changes in the process, and \gls{aoi} is not sufficient to compute the expected \gls{voi}. 

Naturally, we can extend this to multiple dimensions, and different error functions: we can easily imagine a system in which a multidimensional signal $\mathbf{x}(t)$ is reconstructed from multiple noisy measurements from a set of $N$ different sensors, none of which can directly communicate with the others. In this case, computing \gls{voi} is more complex: the receiver has a better picture of the environment, i.e., can receive data from all the sensors, but its knowledge is statistical and partially outdated. On the other hand, each sensor can have relatively precise information about the part of the signal it observes, but a much narrower view of the overall system. Medium access is also a significant issue, and it is generally easier to avoid collisions in pull-based systems, as the receiver can act as a central coordinator and limit the pool of potential transmitters by requesting specific information, or even address nodes directly. A further complication comes from epistemic errors, i.e., events and perturbations not included in the receiver's model of the tracked signal; a purely pull-based system will often take a long time to detect these errors, and operate poorly on incorrect assumptions.

\begin{table}[t]
\centering
\caption{Equivalences: \gls{aoi}, semantic \gls{voi}, and pragmatic \gls{voi}}
\label{tab:equivalences_voi}
\scriptsize
{
\begin{tabular}{l|l|l|ccc}
\toprule
Task & Communication & Process & AoI & VoI (B) & VoI (C) \\ \hline
 & & Martingale & \checkmark & \checkmark & \checkmark \\ \cline{3-6} 
 & \multirow{-2}{*}{Pull-based} & Stateful & & \checkmark & \checkmark \\ \cline{2-6} 
\multirow{-3}{*}{Tracking} & Push-based & Any & & \checkmark & \checkmark \\ \hline
& & Martingale  & \checkmark & \checkmark & \\ \cline{3-6} 
 & \multirow{-2}{*}{Pull-based} & Stateful  & & & \\ \cline{2-6} 
\multirow{-3}{*}{Control} & Push-based & Any &  & &\\
\bottomrule
\end{tabular}}
    \vspace{-0.6cm}
\end{table}

We can then move to the Level C problem by introducing the possibility of control, i.e., of changing the environment to achieve a specific goal \cite{Tung;JSAC:21}. Note that the environment is subject to natural evolution within the physical time, such that the timing requirements depend on the speed of the process relative to the \gls{aoi}, as already indicated for pragmatic communication. In this case, the definition of \gls{voi} can be adapted to consider the performance of the controller instead of the accuracy of the estimator. Unless the control task is simply to track the input signal, the \emph{pragmatic} \gls{voi} is always different from the \emph{semantic} \gls{voi}, as it considers both the content of a sensory observation and its effect on control. There are three possible cases: \emph{(1)}
The update has a low semantic value and does not significantly affect the receiver's estimate of the environment. It can be subject to lossy compression with a limited loss in terms of performance, or even discarded entirely; \emph{(2)} The update has a high semantic value, but it has a low pragmatic value: the change in the receiver's estimate leads to taking the same action, or another action with the same control performance. Thus, a semantic scheduler and compressor would consider this type of update highly important, while it has low relevance to the Level C problem; \emph{(3)} The update significantly affects the receiver's choices, and receiving it would lead to greatly improved control performance; hence, the update has a high pragmatic \gls{voi}, and should be transmitted with a high priority.

The equivalences between \gls{aoi}, semantic \gls{voi}, and pragmatic \gls{voi}, which are Level A, B, and C metrics, respectively, are listed in Table~\ref{tab:equivalences_voi}. 
However, timing metrics such as \gls{aoi} can still be extremely useful as a first approximation of higher-level metrics: in most scenarios, sharp variations are relatively rare, and can be treated as anomalies in a relatively simple weighted \gls{aoi} minimization problem. 

\subsection{Real-time inference over communication links}
\label{RTInference}
Inference at the network edge will empower a broad spectrum of applications, such as industrial automation, extended reality, autonomous vehicles, and robots~\cite{Zhu:COMMAG:2020}. \emph{Edge inference} refers to the process of offloading inference tasks, executed through large-scale \gls{ai} models, from edge devices to edge servers, which are co-located and connected to base stations. 
By leveraging the powerful computational resources of these servers, edge devices can experience enhanced capabilities (e.g., vision, decision making, and natural language processing) and extended battery life. A real-time edge inference can support tactile applications, such as extended reality and robotic control.  The major challenge to this  paradigm is the communication bottleneck resulting from a massive number of mobile devices uploading high-dimensional data features to servers for inference. 

A specific edge inference architecture is \emph{split inference}, which divides a trained model into a low-complexity sub-model on a device and a deep sub-model on a server. The low-complexity sub-model extracts features  from raw data, while the server sub-model performs inference on the uploaded features. This architecture enables resource-constrained edge devices to access large-scale \gls{ai} models on servers while preserving data ownership. 
In split inference, the communication bottleneck is addressed by designing task-oriented techniques, aimed to optimize \gls{e2e} inference throughput, accuracy, or latency. 
To overcome communication constraints, the model's splitting point can be adapted according to available bandwidth and latency requirements \cite{Li:TWC:2020}. For optimizing \gls{e2e} system performance, \gls{jscc} can be employed in split inference to transmit features over the wireless edge. This method follows the same approach as DeepJSCC \cite{Eirina:TCCN:19}, with the  objective of maximizing inference accuracy at the receiver rather than the fidelity of reconstructed features \cite{Jankowski:JSAC:2021}, which is an example of semantic communication. 

\subsubsection{Accelerating Edge Inference by Batching and Early Exiting}

In the field of computing architecture, the well-known von Neumann bottleneck refers to the frequent shuttling of data between memory and processors, which can account for up to 90\% of total computation energy consumption and latency. One technique to address this issue is called batching, which consolidates tasks offloaded by multiple users into a single batch for parallel execution. This process increases the number of tasks executed per unit time by amortizing memory access time and enhancing the utilization of computational resources. Consequently, the latency per task is reduced, facilitating real-time edge inference. However, an excessively large batch size can result in increased queuing time for individual users, leading to reduced throughput.

Another technique that can reduce inference latency is called \emph{early exiting}, which allows a task to exit a \gls{dnn} once it meets the task-specific accuracy requirement. By avoiding the traversal of all network layers uniformly for all tasks, execution speeds are accelerated. Early exiting is characterized by a trade-off between accuracy and computation latency, which is based on a customized \gls{dnn} architecture known as a backbone network \cite{pmlr-v70-bolukbasi17a}. This architecture consists of a conventional \gls{dnn} augmented with multiple low-complexity intermediate classifiers, serving as candidate exit points. A task that requires lower accuracy traverses fewer network layers before exiting, essentially being diverted to an intermediate classifier for immediate inference. Recently, early exiting has been applied to edge inference, specifically in the local/server model partitioning for split inference. Exit points can be jointly optimized to maximize inference accuracy under a latency constraint, while layer skipping and early exiting are combined to facilitate edge inference with stringent resource constraints.

Joint design of batching and early exiting with radio resource management can further enhance \gls{e2e} latency performance, which combines multiple access and inference latency. Specifically, the advantages of end-to-end design are twofold. First, the early-feedback effect, i.e., the instantaneous downloading of inference results upon early exits to intended users, thereby shortening their task latency—is further complemented by batched parallel processing. Second, early exits release computational resources to accelerate other ongoing tasks in the same batch, as reflected by a progressively shrinking batch size over blocks of model layers, which in turn accelerates computation over these blocks. 
However, the complex edge-inference process renders \gls{c2} resource management challenging in at least three ways.  \edit{\emph{(1)}} The interweaving of batching and early exiting, where the latter results in a shrinking batch size over sequential layer blocks, causing variable computation time for different tasks. \edit{\emph{(2)} The task executions of scheduled users have complex interdependencies, not least due to spectrum sharing for communication.}
Specifically, user's accuracy requirements are translated into a variable number of layer blocks to be traversed, while the computation time of each block depends on the random number of tasks passing through it. \edit{\emph{(3)}} The end-to-end latency constraints cause coupling of communication and computation, necessitating joint consideration of channel states and \gls{qos} requirements for batching/scheduling and bandwidth allocation. The optimization of \gls{c2} resource allocation requires solving an integer programming problem that is NP-complete. These challenges present numerous research opportunities in real-time edge inference, such as formulation of the joint optimal control of batching, early exiting, and radio resource management as a combinatorial optimization problem. 

\section{Massive Closed-Loop Communication in 6G}


\edit{Massive communication, in which a very large number of devices communicate with a single base station, is typically associated to the uplink scenario, in which a large number of devices transmit messages to a single receiver. However, as we start to consider semantic and pragmatic communication with a massive number of devices, we move towards a setting in which uplink and downlink are of equal importance and operate in closed-loop. In this section, we will first explain the general role of massive downlink communication in 6G, and then provide an overview of distributed edge inference and learning as an example of massive closed-loop communication relying both on uplink and downlink.}

\subsection{The Role of Massive Downlink Communication}

In 6G applications such as robotic control, distributed learning and edge inference, downlink communication appears as a central feature of massive communication alongside traditional massive uplink. \edit{In some cases, e.g., in distributed learning, the information transmitted in the downlink tends to be the same for a large group of devices and can be realized by a simple broadcast protocol. However, in the case of robotic control or a remote \gls{MDP}, the edge server typically has a specific piece of information that it intends to deliver to a given agent, e.g., based on its own and other agents' states. In current systems, such individual downlink communication often comes at a high cost in terms of overhead, which can be prohibitive when only a small amount of information needs to be communicated (e.g., a robotic action).}

This massive downlink setting with individual information leads to a two-way generalization of the uplink massive random access problem, in which the set of active devices is random. Towards this, we note that the fundamental characteristic of massive communication is the fact that the users are \emph{uncoordinated}. Specifically, instead of considering only massive random access, we can think of \emph{massive uncoordinated communication} as a generalized communication scenario involving both uplink and downlink in which there is high uncertainty about who is active (or relevant) at any given time.

To illustrate the idea and trade-offs for the massive downlink, it is instructive to consider the simple task of communicating 1-bit message acknowledments to a subset of the devices as part of an \gls{arq} feedback scheme. In this case, while the base station knows which devices it successfully received a message from, it does not know the subset of users that it failed to decode, e.g., due to outage. This is in sharp contrast to traditional \gls{arq} in coordinated (scheduled) communication, where the base station knows if it failed to decode a message from a specific user. A consequence of this uncertainty is that the simple task of communicating what used to be a single bit of information, indicating whether the packet was received or not, becomes a highly non-trivial problem. As it turns out, simply concatenating the information for each user with their identifiers is sub-optimal, as illustrated in Fig.~\ref{fig:arq_bounds}~\cite{kalor22arq}. This sub-optimality of concatenation in the massive downlink scenario holds not only for the \gls{arq} problem, but also in the case of transmitting general messages to a small subset of a large population of users~\cite{song23codeddownlink}.

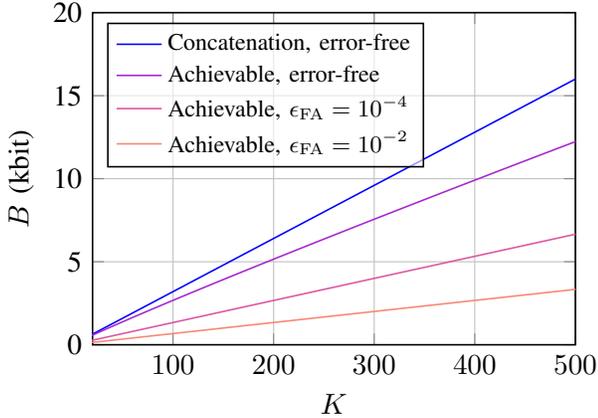
\begin{figure}
    \centering
    \pgfplotsset{scaled y ticks=false}
\begin{tikzpicture}
\begin{axis}[
    xlabel={$K$},
    ylabel={$B$ (kbit)},
    xlabel near ticks,
    ylabel near ticks,
    y label style={at={(axis description cs:-0.1,0.5)},anchor=south},
    xmajorgrids,
    ymajorgrids,
    xmin = 20,
    xmax = 500,
    ytick={0,5000,10000,15000,20000},
    yticklabels={0,5,10,15,20},
    ymin=0,
    ymax=20000,
    width = 8cm,
    height = 6cm,
    semithick,
    legend style={legend cell align=left,font=\footnotesize,legend pos=north west, fill opacity=0.6, draw opacity=1, text opacity=1}]

\addplot+[mark=none,solid,color1] table [col sep=comma,x index=0,y index=1] {fig_tikz/arq_bounds.csv};
\addlegendentry{Concatenation, error-free};

\addplot+[mark=none,solid,color2] table [col sep=comma,x index=0,y index=2] {fig_tikz/arq_bounds.csv};
\addlegendentry{Achievable, error-free};

\addplot+[mark=none,solid,color3] table [col sep=comma,x index=0,y index=4] {fig_tikz/arq_bounds.csv};
\addlegendentry{Achievable, $\epsilon_{\mathrm{FA}}=10^{-4}$};

\addplot+[mark=none,solid,color4] table [col sep=comma,x index=0,y index=3] {fig_tikz/arq_bounds.csv};
\addlegendentry{Achievable, $\epsilon_{\mathrm{FA}}=10^{-2}$};

\end{axis}
\end{tikzpicture}
    \caption{Message length, $B$, required to encode acknowledgment feedback to $K$ out of $2^{32}$ users, either error-free or with false alarm probability $\epsilon_{\mathrm{FA}}$.}
    \label{fig:arq_bounds}
    \vspace{-0.6cm}
\end{figure}

The previous example illustrates the challenge of massive uncoordinated communication at Level A, where the aim is to increase the reliability of the individual messages. Semantic and pragmatic communication point towards alternative ways to make use of massive downlink. For instance, one could construct a semantic \gls{arq} scheme in which the receiver must keep requesting messages until the intended meaning has been correctly received. This requires an acknowledgment message that does not simply inform whether a given message is received or not, but instead tells what is known and unknown to the receiver. The prototypical example could be a surveillance scenario, in which the goal of the receiver is to detect all the objects observed by a number of wireless cameras, e.g., as part of an inference task. By acknowledging the detection of specific objects, as opposed to the individual captured images, a single message can acknowledge all the cameras that capture that specific object. 
\edit{At the pragmatic level, massive downlink is, for instance, relevant for controlling a large fleet of remote \glspl{MDP}, where joint encoding of individual actions might reduce the communication cost at the expense of deviating from the optimal policy. Since the encoding schemes need to be tailored to the specific application and goal, their construction represents an opportunity for future research.}

\subsection{Distributed Inference and Learning at the Edge}
The massive access problem was originally perceived for a massively large number of sensory nodes offloading their occasional bursty measurements in an efficient manner. These measurements are often employed at an edge or cloud server for certain downstream tasks involving inference or learning. However, as the number of sensors and the amount of data collected by them increases, there is a trend towards distributing the learning and inference tasks due to the increasing communication load and privacy concerns. 

6G is likely to give rise to a widespread deployment of edge \gls{ai} to enhance \gls{iot} applications \cite{Zhu:COMMAG:2020} and large-scale distributed sensing enabled by cross-network collaboration among edge devices. The natural integration of edge \gls{ai} and network sensing, known as \gls{aiot} sensing, leverages the advantages of multi-view observations and/or distributed models across many devices, and the powerful prediction capabilities of \gls{dnn} models to enable accurate and intelligent sensing at the edge. This will require combining features/inferences of multiple edge devices transmitted to an edge server over a multiple access channel, or enabling distributed collaborative training across wireless devices.

\edit{Both distributed learning and inference problems require the receiver to carry out computations based on the received features or models. Both scenarios can be considered as examples of semantic communication where the goal is not to convey the underlying features or models reliably, but to enable the receiver to achieve high accuracy in the desired learning task.} Over-the-air computing (AirComp) has emerged as a promising simultaneous-access technique that enables ultra-fast computations over-the-air by leveraging the waveform-superposition property of a multiple access channel \cite{GX2021WCM, Amiri:TSP:20}. The primary motivation behind AirComp is to overcome channel distortion and noise, allowing for accurate implementation of data averaging or other computational functions.  Next we briefly highlight how AirComp can be used to achieve efficient distributed learning and inference in the presence of massive number of devices. 

\subsubsection{Distributed Learning in Massive Access Scenarios}

Federated learning (FL) has emerged as a natural framework for collaborative learning, where an iterative learning algorithm is carried out in a distributed fashion without offloading the data to a central server (see Fig. \ref{f:FL} for an illustration). The devices participating in FL run the necessary iterations locally on their private data, and share only the computed model updates with the \gls{ps} at each iteration. The \gls{ps}, which is the orchestrator of the learning process, aggregates the model updates from the participating devices, and updates the global model, which is then shared with the devices for the next round. This iterative learning process continues for a fixed number of rounds, or until a certain convergence condition is met. 

FL can be implemented across hundreds or even thousands of devices. In certain scenarios, those devices may be in the same physical environment orchestrated by a single \gls{ps}, which may be a macro base station or a WiFi access point, called \gls{feel}. This can be the case for: sensors deployed in the same environment learning a task based on their measurements; mobile devices in the same environment collaboratively learning a communication protocol; or a communication-related task, fine-tuned to the characteristics of the physical environment, e.g.,  millimeter-wave beam alignment based on LIDAR measurements. 

 
 \begin{figure}[t]
     \centering
     \input{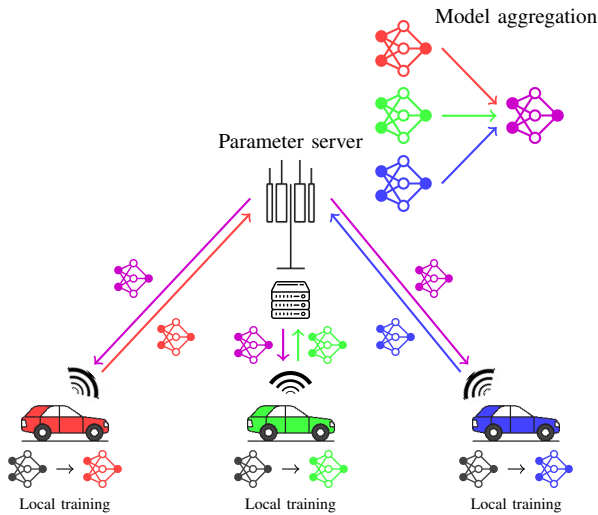}
     \caption{Illustration of \gls{feel}, where the participating devices upload their local updates to the PS through a multiple access channel, while the goal of the PS is to compute their average.}
     \label{f:FL}
    \vspace{-0.6cm}
 \end{figure}

\subsubsection{FEEL with AirComp}
While \gls{feel} with a large number of sensors can be seen as a massive access problem, there is an important distinction. In the classical massive access problem, the goal of the receiver is to decode all the transmitted messages individually, even though the receiver may not be interested in the source of these messages. On the other hand, in the case of FL, the PS does not need individual model updates, as it only needs to recover their averages. 

The computation problem is often treated separately from communication, although this is known to be suboptimal information-theoretically. This was first shown by K\"{o}rner and Marton in \cite{Korner:TIT:79} on a simple example, where the receiver wants to compute the parity of two correlated symmetric binary random variables. They were able to characterize the optimal rate region for this toy example, which was sufficient to illustrate the suboptimality of first sending the sources to the receiver and then carrying out the computation task. Later, Orlitsky showed that the function computation problem is inherently different from conventional source coding by focusing on a point-to-point scenario in \cite{Orlitsky:TIT:01}, where the receiver wants to compute a function of the source available at the transmitter and correlated side information available at the receiver. Furthermore, the optimal rate required for computing any function in a lossless fashion is given by the conditional $G$-entropy, where $G$ is the characteristic graph of the two sources. This is more efficient than first sending the source to the receiver at a rate equal to its conditional entropy given the side information, and then computing the function. 

In the case of \gls{feel}, the goal is to compute the average rather than an arbitrary function. This is a good match for the wireless multiple access, as the  wireless channel superposes the transmitted signals. Averaging is a nomographic function amenable for AirComp. Therefore, in the context of \gls{feel}, wireless interference can be used favorably~\cite{Amiri:TSP:20}, rather than being mitigated, leading to improvement of both the speed and the final accuracy of learning tasks under a bandwidth constraint. 


\subsubsection{FEEL with UMA}

Massive access problem is a generalization of the traditional multiple access problem to a massively large number of transmitters. This formulation adopts the common assumption of the multiple access problem, that the messages are chosen independently at the transmitters, and hence, two simultaneously active users are highly unlikely to transmit the same message. As a consequence, the goal at the receiver at each point in time reduces to estimating the number of active users, and decoding those many messages without necessarily associating them to particular devices. However, when devices are collaborating for a common task, such as distributed learning, their messages can be highly correlated. In this case, the assumption of each transmitter sending a different message is not valid any more, and the receiver must not only decide which messages were transmitted by the active transmitters, but also by how many transmitters. 

For example, in the AirComp solution for the \gls{feel} problem, the model updates are directly modulated to the amplitude of the transmitted symbols. This creates several problems. First, it requires strict synchronisation among devices. 
Second, it assumes continuous amplitude modulation, which is not possible in most practical communication systems, and may lead to problems such as power amplifier saturation or high peak-to-average power ratio. Last, but not least, AirComp requires sending one channel symbol for each model dimension, which corresponds to a very large bandwidth when training large models. 
In \cite{Qiao:ISIT:23}, an alternative digital scheme is proposed for efficient FEEL, based on~\gls{uma} \cite{Polyanskiy:ISIT:17}. Each device first vector quantizes its model update, which is then mapped to a channel codeword. Only a small subset of the devices may be active at each round, but this is not a problem since the PS is not interested in the identity of the active devices, as long as each device is active sufficiently often to contribute to the learning process. Therefore, \gls{uma} techniques can be used here to detect which updates are transmitted by the devices at each round, to be aggregated by the \gls{ps}. However, due to correlation of the model updates, several active devices can transmit the same codeword. Hence, in addition to detecting the active messages, the PS should also count by how many active transmitters each message is transmitted. 

\begin{figure}[t]
\centering
\input{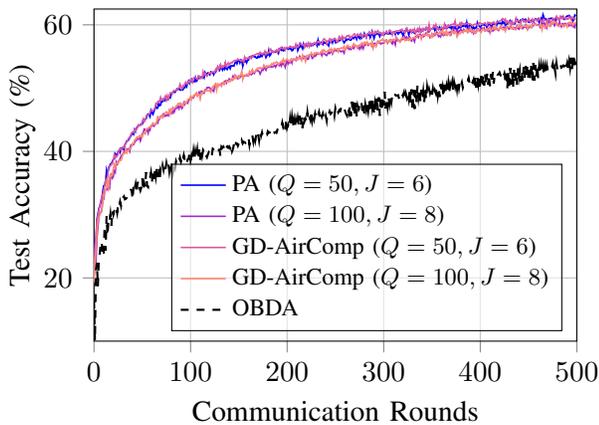}
\caption{Comparison of \gls{gdoac} with PA and OBDA schemes in terms of convergence speed and final test accuracy.}
\label{f:FEEL_UMA}
    \vspace{-0.6cm}
\end{figure}

In Fig. \ref{f:FEEL_UMA}, we plot the convergence behaviour of test accuracy achieved by \gls{pa}, which assumes perfect noise-free channel, \gls{obda} scheme introduced in \cite{GX2021TWC}, and the \gls{uma}-based \gls{gdoac} scheme, proposed above, for the CIFAR-10 image classification task. \edit{Here, $Q$ denotes the size of the compression blocks, where $Q=1$ corresponds to scalar quantization, while $Q=50, 100$ represent vector quantization over a block of symbols, and $J$ refers to $J$-bit quantization; that is, a block of $Q$ symbols is mapped to one of $2^J$ quantization codewords.} For simplicity, we also set the size of the channel codewords transmitted for each block in \gls{gdoac} to $Q$. \edit{OBDA employs $Q=J=1$ with AirComp rather than \gls{uma}.} 
We can observe that \gls{gdoac} not only significantly improves the test accuracy compared to OBDA, but also meets the \gls{pa} baseline in performance.    

\subsubsection{Distributed inference with AirComp}

AirComp can also be used in edge inference to combine features and/or inferences of views/models distributed across multiple devices, which can provide privacy as well as accurate and efficient computation capability \cite{Deniz2022ISIT, Liu:AirPooling:2023}. 

In a multi-view scenario \cite{Su:ICCV:2015}, features extracted from different sensors' views need to be combined into a global feature map, called multi-view pooling, which is then fed into the server's inference model. AirComp-based multi-view pooling is referred to as AirPooling. Implementing average AirPooling using AirComp is relatively straightforward; however, realizing max-pooling with AirComp is considerably more challenging. This is because the class of air-computable functions, known as nomographic functions, are characterized by a summation form with different pre-processing of summation terms and post-processing of the summation \cite{Goldenbaum:TSP:13}. Examples include averaging and geometric mean. The max function in Max AirPooling is not nomographic and does not have a direct AirComp implementation. In  \cite{Liu:AirPooling:2023}, maximization is approximated by using the properties of the generalized p-norm: 
\begin{equation}
        \Vert x\Vert_p=\left(\sum_{n=1}^N |x_n|^p\right)^{\frac{1}{p}}  \begin{cases}
            = \sum_{n=1}^{N} |x_n|, & p = 1, \\
            \rightarrow \max\limits_{n} |x_n|, & p \rightarrow \infty.
        \end{cases}
    \end{equation}
One can see that when the configuration parameter, $p$, is set as one, the function implements averaging; when the parameter \edit{grows to infinity}, the norm approximates maximization. Then a dual-mode AirPooling can be realized by decomposing the air-interface function into pre-processing at devices and post-processing at the server on top of the conventional AirComp and switching the pooling function by controlling $p$.

Consider the popular \gls{e2e} sensing task of classification for object and pattern recognition. The configuration parameter of generated AirPooling should be optimised as it regulates a trade-off between functional approximation and channel-noise suppression. On one hand, the max-function approximation error monotonically decreases as $p$ grows. On the other hand, a larger $p$ tends to amplify channel noise by making the transmitted features, ${x_n}$, highly skewed. Specifically, features with small magnitudes are suppressed and their submission is prone to channel distortion. The main difficulty in the parametric optimisation is the lack of closed-form expression for the \gls{e2e} performance metric, e.g., sensing accuracy. One sub-optimal method as adopted in \cite{Liu:AirPooling:2023} is to use a surrogate metric such as the AirComp error. 


\section{Conclusions and Outlook}

We have presented a view on the wireless 6G systems and its associated technologies by extrapolating the evolution of massive and low-latency connectivity, which were two of the focal points in 5G. This evolution is put in the context of semantic and pragmatic communication in contrast to the technical problem of communication, which was the sole subject of the previous wireless generations. The paper argues that semantic and pragmatic communications have become relevant due to the increasing role that \gls{ml}/\gls{ai} play in communication systems. We presented how the low-latency requirement is expected to evolve towards generalized real-time requirements, coupled with cyber-physical systems and real-time inference. Finally, massive access is considered to evolve towards massive closed-loop communication used in interactive communication with wireless sensors and actuators, as well as in distributed learning. There are other aspects of 6G that were not covered in this article; \edit{for instance, in 3GPP standards there are already efforts for tighter coupling of the medium access layer performance to the real-time goals of the VR/AR traffic.} Nevertheless, the principled discussion of semantic and pragmatic communication is receptive to be expanded with those other 6G aspects, such as integrated communication and sensing or \gls{ntn}.

\bibliographystyle{IEEEtran}
\bibliography{bibliography}

\begin{thebibliography}{10}
\providecommand{\url}[1]{#1}
\csname url@samestyle\endcsname
\providecommand{\newblock}{\relax}
\providecommand{\bibinfo}[2]{#2}
\providecommand{\BIBentrySTDinterwordspacing}{\spaceskip=0pt\relax}
\providecommand{\BIBentryALTinterwordstretchfactor}{4}
\providecommand{\BIBentryALTinterwordspacing}{\spaceskip=\fontdimen2\font plus
\BIBentryALTinterwordstretchfactor\fontdimen3\font minus
  \fontdimen4\font\relax}
\providecommand{\BIBforeignlanguage}[2]{{%
\expandafter\ifx\csname l@#1\endcsname\relax
\typeout{** WARNING: IEEEtran.bst: No hyphenation pattern has been}%
\typeout{** loaded for the language `#1'. Using the pattern for}%
\typeout{** the default language instead.}%
\else
\language=\csname l@#1\endcsname
\fi
#2}}
\providecommand{\BIBdecl}{\relax}
\BIBdecl

\bibitem{popovski2020semantic}
P.~Popovski, O.~Simeone, F.~Boccardi, D.~G{\"u}nd{\"u}z, and O.~Sahin,
  ``Semantic-effectiveness filtering and control for post-5g wireless
  connectivity,'' \emph{J. Indian Inst. Sci.}, vol. 100, pp. 435--443, 2020.

\bibitem{lan2021semantic}
Q.~Lan, D.~Wen, Z.~Zhang, Q.~Zeng, X.~Chen, P.~Popovski, and K.~Huang, ``What
  is semantic communication? a view on conveying meaning in the era of machine
  intelligence,'' \emph{J. Comm. Inf. Netw.}, vol.~6, no.~4, pp. 336--371,
  2021.

\bibitem{gunduz2022beyond}
D.~G{\"u}nd{\"u}z, Z.~Qin, I.~E. Aguerri, H.~S. Dhillon, Z.~Yang, A.~Yener,
  K.~K. Wong, and C.-B. Chae, ``Beyond transmitting bits: Context, semantics,
  and task-oriented communications,'' \emph{IEEE J. Sel. Areas Comm.}, vol.~41,
  no.~1, pp. 5--41, 2022.

\bibitem{popovski2022perspective}
P.~Popovski, F.~Chiariotti, K.~Huang, A.~E. Kal{\o}r, M.~Kountouris, N.~Pappas,
  and B.~Soret, ``A perspective on time toward wireless {6G},'' \emph{Proc.
  IEEE}, vol. 110, no.~8, pp. 1116--1146, 2022.

\bibitem{ShannonWeaver49}
C.~E. Shannon and W.~Weaver, \emph{The Mathematical Theory of
  Communication}.\hskip 1em plus 0.5em minus 0.4em\relax Urbana, IL: University
  of Illinois Press, 1949.

\bibitem{gleason2022development}
J.~B. Gleason and N.~B. Ratner, \emph{The Development of Language}.\hskip 1em
  plus 0.5em minus 0.4em\relax Plural Publishing, 2022.

\bibitem{Dobrushin}
R.~Dobrushin and B.~Tsybakov, ``Information transmission with additional
  noise,'' \emph{IRE Trans. Inf. Theory}, vol.~8, no.~5, pp. 293--304, 1962.

\bibitem{Eirina:TCCN:19}
E.~{Bourtsoulatze}, D.~B. {Kurka}, and D.~{G\"und\"uz}, ``Deep joint
  source-channel coding for wireless image transmission,'' \emph{IEEE Trans.
  Cogn. Comm. Netw.}, vol.~5, no.~3, pp. 567--579, 2019.

\bibitem{Kurka:TWC:21}
D.~B. Kurka and D.~G\"und\"uz, ``Bandwidth-agile image transmission with deep
  joint source-channel coding,'' \emph{IEEE Trans. Wireless Comm.}, vol.~20,
  no.~12, pp. 8081--8095, 2021.

\bibitem{Tung;JSAC:21}
T.-Y. Tung, S.~Kobus, J.~P. Roig, and D.~Gündüz, ``Effective communications:
  A joint learning and communication framework for multi-agent reinforcement
  learning over noisy channels,'' \emph{IEEE J. Sel. Areas Comm.}, vol.~39,
  no.~8, pp. 2590--2603, 2021.

\bibitem{Kobus:ISIT:23}
S.~Kobus, T.-Y. Tung, and D.~Gündüz, ``Goal-oriented compression with a
  constrained decoder,'' in \emph{Proc. IEEE Int. Symp. Inf. Theory (ISIT)},
  2023.

\bibitem{alawad2022value}
F.~Alawad and F.~A. Kraemer, ``Value of information in wireless sensor network
  applications and the {IoT}: A review,'' \emph{IEEE Sensors J.}, vol.~22,
  no.~10, pp. 9228--9245, 2022.

\bibitem{holm2023goal}
J.~Holm, F.~Chiariotti, A.~E. Kal{\o}r, B.~Soret, T.~B. Pedersen, and
  P.~Popovski, ``Goal-oriented scheduling in sensor networks with application
  timing awareness,'' \emph{IEEE Trans. Comm.}, vol.~71, pp. 4513--4527, 2023.

\bibitem{Zhu:COMMAG:2020}
G.~Zhu, D.~Liu, Y.~Du, C.~You, J.~Zhang, and K.~Huang, ``Toward an intelligent
  edge: Wireless communication meets machine learning,'' \emph{IEEE Comm.
  Mag.}, vol.~58, no.~1, p. 19–25, 2020.

\bibitem{Li:TWC:2020}
E.~Li, L.~Zeng, Z.~Zhou, and X.~Chen, ``Edge {AI}: On-demand accelerating deep
  neural network inference via edge computing,'' \emph{IEEE Trans. Wireless
  Comm.}, vol.~19, no.~1, p. 447–457, 2020.

\bibitem{Jankowski:JSAC:2021}
M.~Jankowski, D.~Gündüz, and K.~Mikolajczyk, ``Wireless image retrieval at
  the edge,'' \emph{IEEE J. Sel. Areas Comm.}, vol.~39, no.~1, pp. 89--100,
  2021.

\bibitem{pmlr-v70-bolukbasi17a}
T.~Bolukbasi, J.~Wang, O.~Dekel, and V.~Saligrama, ``Adaptive neural networks
  for efficient inference,'' in \emph{Proc. Int. Conf. Mach. Learn. (ICML)},
  2017.

\bibitem{kalor22arq}
A.~E. Kal{\o}r, R.~Kotaba, and P.~Popovski, ``Common message acknowledgments:
  Massive {ARQ} protocols for wireless access,'' \emph{IEEE Trans. Comm.},
  vol.~70, no.~8, pp. 5258--5270, 2022.

\bibitem{song23codeddownlink}
R.~Song, K.~M. Attiah, and W.~Yu, ``Coded downlink massive random access,'' in
  \emph{Proc. IEEE Int. Symp. Inf. Theory (ISIT)}, 2023.

\bibitem{GX2021WCM}
G.~Zhu, J.~Xu, K.~Huang, and S.~Cui, ``Over-the-air computing for wireless data
  aggregation in massive {I}o{T},'' \emph{IEEE Wireless Comm.}, vol.~28, no.~4,
  pp. 57--65, 2021.

\bibitem{Amiri:TSP:20}
M.~{M. Amiri} and D.~{G\"und\"uz}, ``Machine learning at the wireless edge:
  {D}istributed stochastic gradient descent over-the-air,'' \emph{IEEE Trans.
  Signal Process.}, vol.~68, pp. 2155--2169, 2020.

\bibitem{Korner:TIT:79}
J.~K\"{o}rner and K.~Marton, ``How to encode the modulo-two sum of binary
  sources (corresp.),'' \emph{IEEE Trans. Inf. Theory}, vol.~25, no.~2, pp.
  219--221, 1979.

\bibitem{Orlitsky:TIT:01}
A.~Orlitsky and J.~Roche, ``Coding for computing,'' \emph{IEEE Trans. Inf.
  Theory}, vol.~47, no.~3, pp. 903--917, 2001.

\bibitem{Qiao:ISIT:23}
L.~Qiao, Z.~Gao, Z.~Li, and D.~G\"und\"uz, ``Unsourced massive access-based
  digital over-the-air computation for efficient federated edge learning,'' in
  \emph{Proc. IEEE Int. Symp. Inf. Theory (ISIT)}, 2023.

\bibitem{Polyanskiy:ISIT:17}
Y.~Polyanskiy, ``A perspective on massive random-access,'' in \emph{Proc. IEEE
  Int. Symp. Inf. Theory (ISIT)}, 2017, pp. 2523--2527.

\bibitem{GX2021TWC}
G.~Zhu, Y.~Du, D.~Gündüz, and K.~Huang, ``One-bit over-the-air aggregation
  for communication-efficient federated edge learning: Design and convergence
  analysis,'' \emph{IEEE Trans. Wireless Comm.}, vol.~20, no.~3, pp.
  2120--2135, 2021.

\bibitem{Deniz2022ISIT}
S.~F. Yilmaz, B.~Hasırcıoğlu, and D.~Gündüz, ``Over-the-air ensemble
  inference with model privacy,'' in \emph{Proc. IEEE Int. Symp. Inf. Theory
  (ISIT)}, 2022.

\bibitem{Liu:AirPooling:2023}
Z.~Liu, Q.~Lan, A.~E. Kal{\o}r, P.~Popovski, and K.~Huang, ``Over-the-air
  multi-view pooling for distributed sensing,'' \emph{arXiv preprint
  arXiv:2302.09771}, 2023.

\bibitem{Su:ICCV:2015}
H.~Su, S.~Maji, E.~Kalogerakis, and E.~Learned-Miller, ``Multi-view
  convolutional neural networks for {3D} shape recognition,'' \emph{Proc. IEEE
  Int. Conf. Comput. Vis. (ICCV)}, 2015.

\bibitem{Goldenbaum:TSP:13}
M.~Goldenbaum, H.~Boche, and S.~Stańczak, ``Harnessing interference for analog
  function computation in wireless sensor networks,'' \emph{IEEE Trans. Signal
  Process.}, vol.~61, no.~20, pp. 4893--4906, 2013.

\end{thebibliography}

\end{document}